%% file: pLaPaper.tex
\documentclass[aps,prc,twocolumn,preprintnumbers,superscriptaddress,showpacs,showkeys]{revtex4}

\usepackage{graphicx}
\usepackage{amsmath}

\newcommand{\sqrtsNN}{\mbox{$\sqrt{\mathrm{s}_{_{\mathrm{NN}}}}$}}
\newcommand{\alam}{\overline{\Lambda}}
\newcommand{\axiz}{\overline{\Xi^{0}}}
\newcommand{\axim}{\overline{\Xi^{-}}}
\newcommand{\asigz}{\overline{\Sigma^{0}}}
\newcommand{\lam}{\Lambda}
\newcommand{\kStar}{k^{*}}
\newcommand{\Kz}{K^{0}_{S}}
\newcommand{\pbar}{\overline{p}}
\newcommand{\dedx}{dE/dx}
\newcommand{\pt}{p_{t}}
\newcommand{\xim}{\Xi^-}
\newcommand{\xiz}{\Xi^0}
\newcommand{\sigz}{\Sigma^{0}}
\newcommand{\sigp}{\Sigma^{+}}
\newcommand{\asigp}{\overline{\Sigma^{+}}}

\begin{document}

\title{Proton - $\Lambda$ correlations in central Au+Au collisions at \sqrtsNN\ =~200~GeV \\}

\include{STAR_authors}

\date{\today}

\begin{abstract}
We report on $p-\lam$, $p-\alam$, $\pbar-\lam$ and $\pbar-\alam$ correlation functions
constructed in central Au-Au collisions at \sqrtsNN = 200 GeV by the STAR experiment
at RHIC. The proton and lambda source size is
inferred from the $p-\lam$ and $\pbar-\alam$ correlation functions.
It is found to be smaller than the pion source size also
measured by the STAR experiment
at smaller transverse masses, in agreement with a scenario
of a strong universal collective flow.
The $p-\alam$ and $\pbar-\lam$
correlation functions,
which are measured for the first time, exhibit a large anti-correlation.
Annihilation channels and/or a negative real part of the spin-averaged
scattering length must be included in the final-state interactions
calculation to reproduce the measured correlation function.
\end{abstract}



\maketitle

\section{INTRODUCTION}
Correlations amongst non-identical particles are sensitive to the
space-time extent of their emitting source
(see e.g. \cite{Led:93B373}).
Originally uncorrelated particles produced in nearby phase space points
in the
prompt
emission final state can interact through
the nuclear and/or the Coulomb force
and become correlated
at time scales much longer
than the production time. When the final-state
interaction (FSI) is relatively well understood the emitting
source size can be inferred from correlations at small relative
three-velocity of the particles in their center-of-mass system.
In relativistic heavy-ion collisions large particle
densities are produced and the collision fireball may undergo a
collective expansion (i.e. \ flow) \cite{Adams:2004c,Adams:2004a}.
This flow can induce
space-momentum correlation so that particles
with similar velocities
come from the near-by regions of the source.
With a strong flow at RHIC,
as suggested by
several measurements (see e.g. \cite{Adams:2004b,Adams:2004c,Adams:2004a,Adams:2004e}),
the observed source sizes should be reduced
relative to a source without flow \cite{BW:2003} and
vary with the mass of the emitted
particle:
the heavier the particle, the smaller
is the reduction of the collective
flow effect due to the thermal motion
and the smaller are the apparent
source sizes
\cite{Adams:2004e,Blume:2003}.
This flow effect can also be studied with $p-p$
correlations and compared with
$\pi-\pi$ or $K-K$ correlations.
As compared with the $p-p$ system,
the $p-\lam$ system gains in
statistics in the region of small relative
velocities due to the absence of
repulsive Coulomb interaction \cite{Wang:9983}.

In this paper we test the hypothesis that a strong flow is established
in Au+Au collisions at \sqrtsNN = 200 GeV by comparing the source
sizes of protons and lambdas to that of pions. The first measurements
of $\pbar-\lam$, $p-\alam$ and $\pbar-\alam$ correlation
functions are presented.
The $p-\lam$ and $\pbar-\alam$
interaction potentials
are relatively well understood,
so we are able to infer source
sizes \cite{Blume:2003,Chung:2003,Wang:9983,Rijnucl:1998,Maessen:1989,Nagels:1979}.
The $\pbar-\lam$ and $p-\alam$ FSI, on the other hand, are
unknown. As such, the scattering lengths and source sizes
are extracted by fitting
the data from the STAR experiment with
the Lednick\'y \& Lyuboshitz analytical model \cite{Led:90COR}.
Besides constraining baryon - antibaryon potentials,
this information determines unknown $\pbar-\lam$ and
$p-\alam$ annihilation cross sections which are useful
to constrain heavy-ion cascade models \cite{Wang:a9807}.

\section{DATA RECONSTRUCTION}

\subsection{Events selection}
The analysis was carried out using the STAR detector
at RHIC \cite{Ackermann:1999a}.
Two million Au+Au collisions have been analyzed with \sqrtsNN = 200 GeV.
Because of statistics issues, only the
$10\%$ most central collisions were selected with
the Zero-Degree Calorimeters and the Central Trigger Barrel
of the STAR detector. This event selection procedure is
explained in detailed in \cite{Adams:2004e}.
The other centrality selections didn't gave any
statistically meaningful results.
Tracking of charged particles was accomplished using
the STAR Time Projection Chamber (TPC)
which covers the kinematic range of transverse
momentum $\pt >150~\mathrm{MeV}/c$,
pseudo-rapidity $|\eta|<1.5$
and azimuthal angle $0<\phi<2\pi$.
Events analyzed in this paper have collision
vertices within $\pm25~\mathrm{cm}$
longitudinally of the TPC center.

\subsection{Protons and antiprotons selection}
Protons and antiprotons
are identified using their specific energy loss ($\dedx$)
in the TPC gas.
This selection limits the acceptance of particles to
the transverse momentum range of $0.4-1.1~\mathrm{GeV/c}$ in the
rapidity interval $|y|<0.5$.
Tracks pointing to within $3~\mathrm{cm}$ of the primary vertex are included
in the primary track sample.

\subsection{Lambdas and antilambdas selection}
\begin{figure}[!ht]
\center
\includegraphics[width=9cm, height=7.5cm]{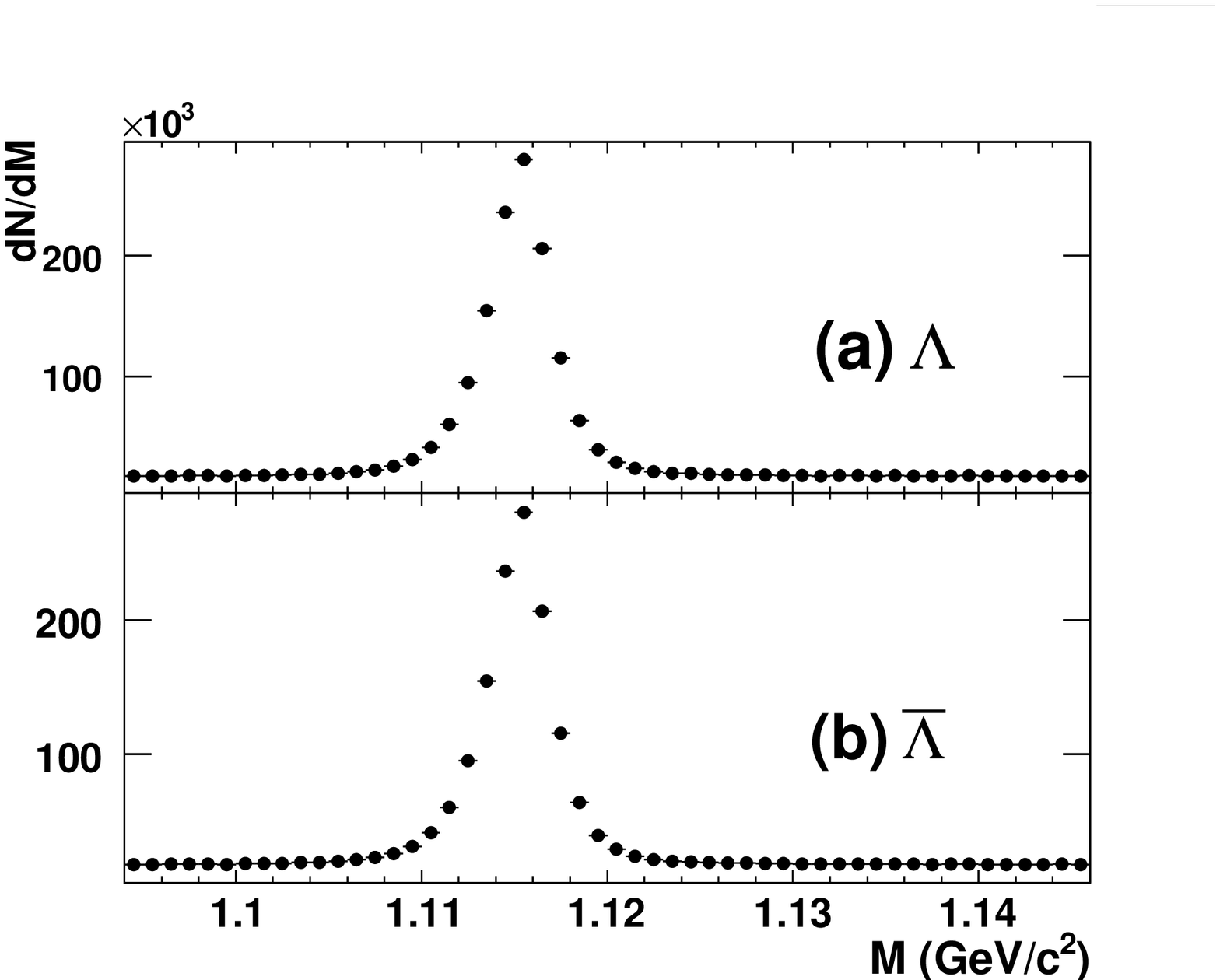}
\caption{(Color online) Invariant mass of the selected $\lam$ (a) and $\alam$ (b) background not substracted with $0.3<\pt<2.0~\mathrm{GeV/c}$. The y-axis represents the number of candidates used in this analysis.}
\label{minv}
\end{figure}

Lambdas (antilambdas) are reconstructed through the decay channel
$\Lambda \rightarrow \pi^{-} + p$
($\overline{\Lambda} \rightarrow \pi^{+} + \overline{p}$) \cite{Adams:2003},
with branching ratio of $64\%$.
Pions and protons (i.e. lambda daughters) are selected using their
specific energy loss.
The invariant mass (Figure \ref{minv}) range of the lambda candidates is $1115\pm6~\mathrm{MeV/c^{2}}$, the $\pm6~\mathrm{MeV/c^{2}}$ has been fixed to optimize the signal over noise ratio.
The signal over noise ratio is equal to $86\pm6\%$
for lambda ($\langle \pt \rangle = 1.05~\mathrm{GeV/c}$)
and for antilambda ($\langle \pt \rangle = 1.09~\mathrm{GeV/c}$)
in the $\pm6~\mathrm{MeV/c^{2}}$ mass window and $0.3<\pt<2.0~\mathrm{GeV/c}$.
The correlation effect was the same within the errors with one sigma
cut ($\pm3~\mathrm{MeV/c^{2}}$) on the invariant mass.
In addition the following geometrical cuts
are applied.
The distance of the closest approach (DCA) of lambda daughters
is required to be less than $0.7~\mathrm{cm}$.
The DCA of the decay pions with respect to the primary vertex
is required to be greater than $2.0~\mathrm{cm}$.
The DCA of the reconstructed neutral particles to the primary vertex
is required to be less than $0.6~\mathrm{cm}$.
To avoid $\Kz$ being misidentified as lambdas, lambda candidates are
rejected if their invariant mass is within the window
$497.7^{+10.0}_{-21.3}~\mathrm{MeV/c^{2}}$
when the pion mass is assumed for the two daughters.
Due to the detector acceptance and the selection criteria, the $\pt$
range of the lambda sample is $0.3<\pt<2.0~\mathrm{GeV/c}$
and $|y|<1.5$.

\subsection{Pairs selection}
When studying two-article FSI,
the relevant variable is the momentum of one
of the particles in the pair rest frame called here
$\vec{k}^*$ ($\kStar = |\vec{k}^*|$).
The correlation function has been extracted by
constructing the ratio of two distributions.
The numerator is the $\kStar$ distribution
of pairs from the same event.
The denominator
is the $\kStar$ distribution of pairs composed
of particles from different events
with primary vertices separated from each other
by less than $10~\mathrm{cm}$.
The ratio is formed by dividing the numerator by the denominator. Then
the ratio is normalized to 1 at high $\kStar$ ($\kStar > 0.35~\mathrm{GeV/c}$).
The event mixing procedure
is the same as the one used in \cite{Adams:2004e}.
When reconstructing a primary lambda (antilambda)
the decay proton (antiproton) points directly
back to the primary vertex
and may share some hits with a
primary proton (antiproton) in the TPC.
This phenomenon is called
track merging
and can occur while building pairs for
the correlation function.
In case of track merging, instead of counting two tracks with small $\kStar$
only one track will be found. So one pair will be missed at small $\kStar$.
If a lot of pairs are missed, the correlation function will show a hole at
small $\kStar$ because they are not found.
A missed pair leads to a fake correlation because of the event mixing
procedure. A pair can be missed in a real event (in the numerator).
Such a pair may be reconstructed taking two different events in order to
build the background (in the denominator).
Thus, track merging leads to fake correlations.
For two tracks of different
momenta, or different polar angles,
the number of shared hits may vary as a function of where they cross
in the TPC. It could be as low as 5 hits on the edge of the TPC. The
tracker can not find these hits, it is linked with the finding seed.
This affects high $\pt$ tracks because they will have more hits merged.
To avoid such fake correlations, track merging has
been studied for all other
possible track - daughter track combinations.
The study of track merging for
lambda daughters and
proton/antiproton tracks
leads to two different selections criteria.
The first selection requires tracks to share fewer
than $10\%$ of their TPC space points.
The second selection deals with
the average separation between primary tracks
and lambda/antilambda daughter tracks.
The track separation is calculated as an arithmetic mean
distance between the TPC hits of the two tracks for a given radius.
If a track crosses the whole TPC, it will be reconstructed
with a maximum of 45 hits.
Because one of the lambda/antilambda daughter
tracks is a secondary track,
all 45 hits of the TPC may not be available,
so the mean is calculated from a maximum of 11 distances.
In this paper "secondary particles" means particle from decay.
As a consequence the average separation between
protons/antiprotons and lambda/antilambda daughters
are required to be greater than
$11~\mathrm{cm}$ for $\pbar-\pbar_{\alam}$,
$\pbar-p_{\lam}$, $\pbar-\pi_{\lam}$, $p-\pi_{\alam}$,
$10~\mathrm{cm}$ for $p-p_{\lam}$, $12~\mathrm{cm}$ for $\pbar-\pbar_{\alam}$,
$17~\mathrm{cm}$ for $p-\pi_{\lam}$ and $\pbar-\pi_{\alam}$.
The first selection prevents interference between opposite sign tracks
that, even though their average separation is large, can cross each
other in the TPC. When the trajectories of the particles cross, space
points can be assigned to the wrong track during reconstruction.
In some cases this can lead to a failure to reconstruct one of the tracks.
For this reason the values for the minimum average separation are
larger for opposite sign pairs.


\section{PURITY}

\subsection{Definition}
Impurities in the sample of protons and lambdas will reduce
the observed $p-\lam$ correlation strength. In the case of lambdas,
fake lambda candidates (from combinatorial background) and secondary
lambdas (e.g. from $\sigz$ decays) are the two
main sources of impurity. The sample of protons
is contaminated by other charged tracks falsely identified as protons
and by
protons from weak decays (feed-down). In order to correct the observed
correlations for misidentification and feed-down we estimate the
particle purity for $p$, $\pbar$, $\lam$ and $\alam$ as a function of
transverse momentum ($\pt$):
\begin{eqnarray}
\label{Purity}
\textrm{Particle Purity} (\pt) = \textrm{Pid} (\pt) \times \textrm{Fp} (\pt),
\end{eqnarray}
where Pid is \textit{the probability a candidate was correctly identified} and
Fp is \textit{the fraction of the candidates that were primary particles}.
The final correction depends on the product
of the particle purity for both
particles (i.e. the Pair Purity).
The pair inpurity is
corrected for in constructing the correlation function in $\kStar$.

The feed-down estimations have been done for
$p$, $\pbar$, $\lam$ and $\alam$ and are
summarised in Table \ref{Tabpurity}.
Combined results from STAR \cite{Adler:2002a,Lamont:2002th,Adams:2004c,Adams:2004a} and
predictions from a
thermal model
\cite{Braun-Munzinger:1996A606,Braun-Munzinger:1999B465,Braun-Munzinger:2001ip}
have been used.
The approximations introduced by estimating
the purity are the major
source of systematic uncertainties
on the extracted values of FSI
parameters and source radii discussed below.

\subsection{Protons and antiprotons purities}
The identification probabilities have been estimated
for charged particles;
they are also given in Table \ref{Tabpurity}.
A track can be identified as
a pion, a kaon, a proton/antiproton
and an electron/positron
with a certain probability
using the information about
the energy loss $\dedx$ \cite{Adler:2001a}.
Identified protons (antiprotons) from the selected sample
have a mean identification probability of $76\pm7\%$ ($74\pm7\%$).
This mean identification probability and the corresponding $\pt$ are
calculated over
all selected tracks considered as protons (antiprotons).
The calculated feed-down
leads to a mean estimated fraction of primary protons
of $52\%$
(with a mean transverse velocity
$\langle \beta_t \rangle \equiv \langle p_t / \gamma \rangle / m ) = 0.58$,
$m$ is the mass of the particle).
Most of the secondary protons come
from lambda decays
(primary $\lam$ and $\lam$ from $\sigz$, $\xiz$ and $\xim$)
and constitute $36\%$ of the protons
used to construct the correlation function.
Other sources of contamination for
protons are products of $\Sigma^{+}$
decays and interactions of pions with detector materials
which represent respectively
$10\%$ and $2\%$ of the sample.
The feed-down study for antiprotons
($\langle \beta_t \rangle=0.60$),
leads to an estimated fraction of primary antiprotons of $48\%$.
Most of the secondary antiprotons come
from antilambda decay (primary $\alam$ and $\alam$
from $\asigz$, $\axiz$ and $\axim$)
and constitute $39\%$ of the antiprotons
used to construct the correlation function.
Antiprotons from $\overline{\Sigma^{+}}$ decays
are another major source of
contamination; they make up $13\%$ of the antiproton sample.

\subsection{Lambdas and antilambdas purities}
For lambdas and antilambdas the probability of
misidentification corresponds to background estimation
under the mass peak.
The corresponding identification probabilities
are practically independent of $\pt$ and equal
to $86\pm6\%$ for both
lambdas and antilambdas, respectively.
The sample of lambdas (antilambdas) includes secondary particles
such as decay products of
$\xim, \xiz, \sigz$
($\axim, \axiz, \asigz$).
The fractions of primary lambdas
($\langle \beta_t \rangle =0.68$)
and primary antilambdas
($\langle \beta_t \rangle=0.70$)
have been estimated at $45\%$.

\subsection{Pairs purities}

\begin{table}[hbt!]
\begin{tabular}{|c|c|c|c|}
 \hline
Particle & Identification & Fraction Primary \\ \hline
$p$     & $76\pm7\%$    &       $52\pm4\%$\\ \hline
$\pbar$ & $74\pm7\%$    &       $48\pm4\%$\\ \hline
$\lam$  & $86\pm6\%$    &       $45\pm4\%$\\ \hline
$\alam$ & $86\pm6\%$    &       $45\pm4\%$\\ \hline
\end{tabular}
\caption{Summary of the particle purity due to
identifications and weak-decay contamination.
Values are averaged over the transverse momentum
without taking into account the transverse momentum
dependence for $\kStar < 0.2~\mathrm{GeV/c}$.}
\label{Tabpurity}
\end{table}

The pair purity
plays a crucial role in the correlation study.
The estimated value of the mean pair purity
for $p-\lam$,
$\pbar-\alam$, $p-\alam$ and
$\pbar-\lam$ systems is $\lambda = 17.5\pm2.5\% $ after taking into
account the transverse momentum dependence.
Without taking
into account the transverse momentum dependence
the estimated purities
differ by $2\%$ (Tab. \ref{Tabpurity}).

\section{CORRECTIONS}

\subsection{Purity}
Because the contamination reduces the correlation strength,
raw data have been corrected for purity using the relation:
\begin{eqnarray}
\label{Ctrue}
C_{measured}^{corr}(k^*) = \frac{C_{measured}(k^*)-1}{\textrm{PairPurity}} + 1,
\end{eqnarray}
where PairPurity is the product of the purities for the two particles
and $C_{measured}^{corr}(k^*)$ and $C_{measured}(k^*)$
are respectively the corrected and measured correlation functions.
Eq. (\ref{Ctrue}) assumes that misidentified and weak decay protons (antiprotons)
are uncorrelated with lambdas and antilambdas.
This assumption is justified for misidentified protons (antiprotons)
since the eventual pion or kaon correlation at small $\kStar$ is washed out
after the wrong mass assignment. Combinatoric background reconstructed as
$\lam$ and $\alam$ also leads to uncorrelated pairs. On the other hand weak
decay products may keep a residual correlation from their parents. This
assumption will be revisited when extracting source sizes and
scattering lengths from the correlation functions.

\subsection{Momentum resolution}
The effects of momentum resolution have been studied
using mixed pairs by calculating the weight
with the Lednick\'y \& Lyuboshitz analytical model \cite{Led:90COR}.
It appears
that compared with statistical and systematic errors,
the impact of the momentum resolution effect is negligible.
Indeed, the momentum resolution effect leads to about
$1\%$ variation of the apparent source radius.
Nevertheless, correlation functions have been corrected
for the momentum resolution using the following formula:
\begin{eqnarray}
C_{true}(k^*) = \frac{C_{measured}^{corr}(k^*) \times C_{Th-not-smeared}(k^*)}{C_{Th-smeared}(k^*)},
\end{eqnarray}
where $C_{true}(k^*)$ represents the corrected correlation function,
$C_{Th-not-smeared}(k^*)/C_{Th-smeared}(k^*)$ is the correction factor;
$C_{Th-not-smeared}(k^*)$ is calculated without taking into account
the effect of momentum resolution and $C_{Th-smeared}(k^*)$
includes this effect. The shift due to the momentum resolution is studied
using simulated tracks introduced into real events. This shift is applied
to momenta in order to calculate $C_{Th-smeared}(k^*)$.

\begin{figure}[!ht]
\center
\includegraphics[width=9cm, height=7.5cm]{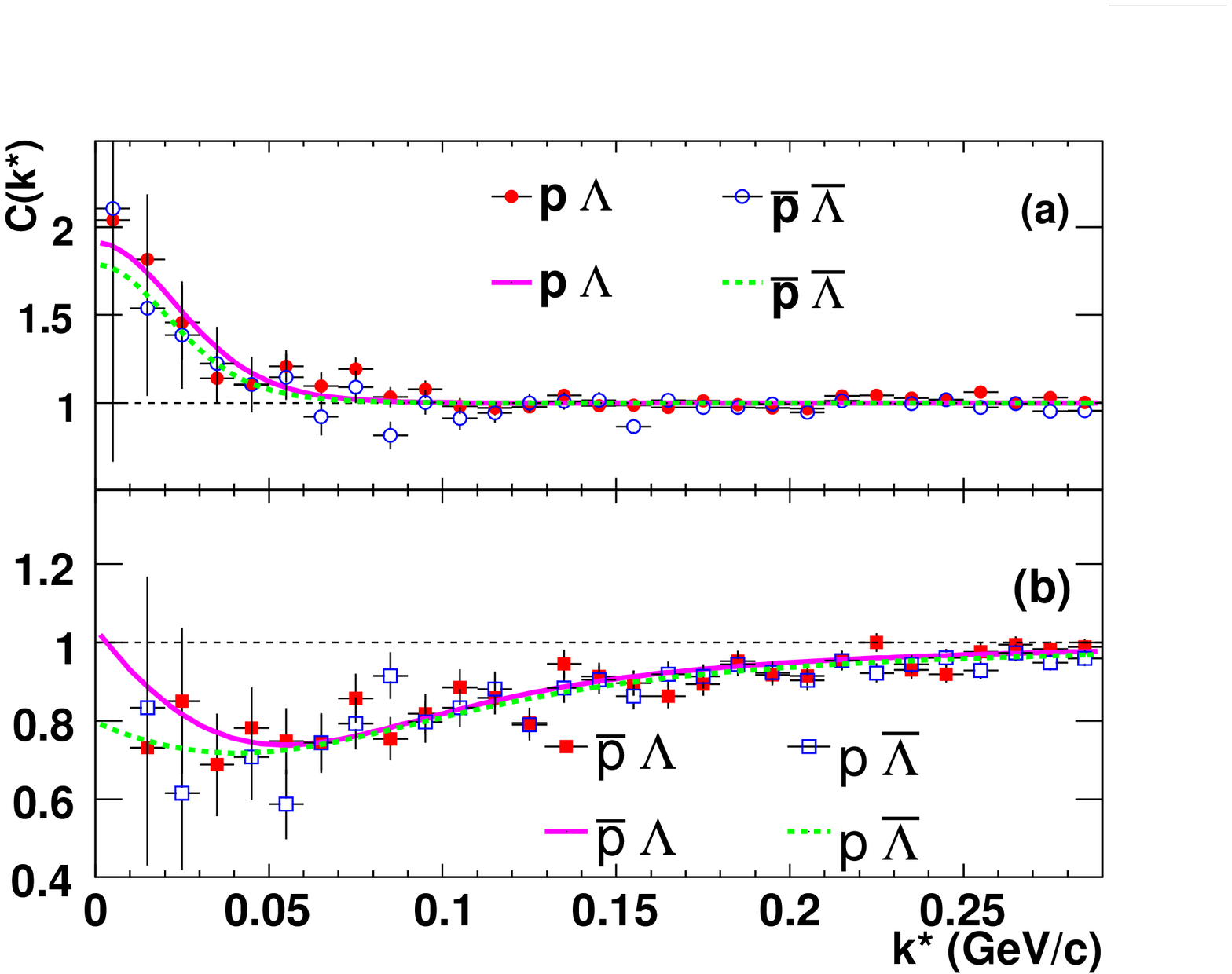}
\caption{(Color online) The purity and momentum-resolution
corrected correlation functions $C_{true}(k^*)$ for
$p-\lam$, $\pbar-\alam$ (a), $\pbar-\lam$, $p-\alam$ (b).
Curves correspond to fits done using the
Lednick\'y \& Lyuboshitz analytical model \cite{Led:90COR}.}
\label{plaCorrFit}
\end{figure}

\section{RESULTS}

\subsection{Correlation functions}
\begin{figure}[htb!]
\center
\includegraphics[width=9cm, height=7.5cm]{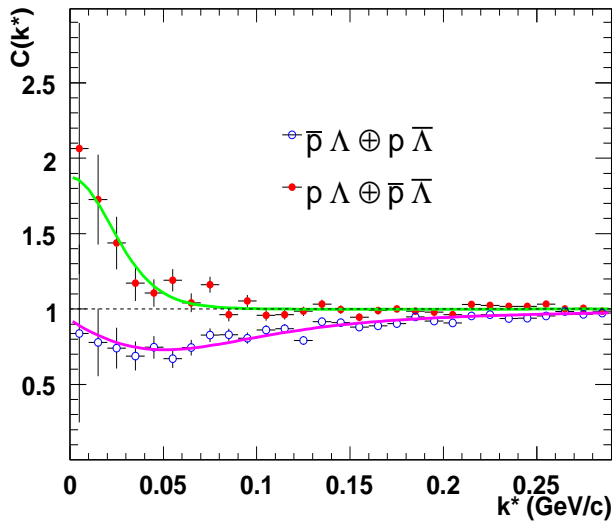}
\caption{(Color online) ($p-\lam)\oplus(\pbar-\alam)$ and
$(\pbar-\lam)\oplus(p-\alam)$ combined correlation functions.
Correlation functions are corrected for purity and momentum resolution.
Curves correspond to fits done using
the Lednick\'y \& Lyuboshitz analytical model \cite{Led:90COR}.}
\label{plaComb}
\end{figure}

In Fig.\ref{plaCorrFit} (a) the corrected $p-\lam$ and $\pbar-\alam$
correlation functions are shown. They are close to each other,
within error bars,
showing a pair excess at
small $k^{*}$ ($0 < k^{*} < 0.1~\mathrm{GeV/c}$)
indicating an attractive potential
between (anti)proton and
(anti)lambda.
Fig.\ref{plaCorrFit} (b) shows
the corrected $\pbar-\lam$ and $p-\alam$ correlation functions
measured for the first time.
They are below unity in a wide $k^{*}$ range
$0 < k^{*} < 0.25~\mathrm{GeV/c}$
consistent with positive imaginary parts of the s-wave scattering
lengths (due to the open annihilation channels) and a negative
real part of the spin averaged s-wave scattering length.
In Figure \ref{plaComb} the combined
$(p-\lam)\oplus (\pbar-\alam)$ and
$(\pbar-\lam)\oplus(p-\alam)$ correlation functions
are presented. The symbol $\oplus$ means that
numerators and denominators of the systems
have been added to build the combined correlation
functions.
In both Figures, curves correspond to a fit
carried out with the Lednick\'y \& Lyuboshitz analytical model \cite{Led:90COR}.

\subsection{Lednick\'y \& Lyuboshitz analytical model}
This model relates
the two-particle correlation functions
with source sizes and scattering amplitudes \cite{Led:01Pal,Led:90COR}.
As usual, similar to the Fermi factor in the
theory of $\beta$-decay, the
correlation function ($C(\kStar)$)
is calculated as the square of the wave function ($\Psi^{S}$)
averaged over the total spin S and over
the distribution of relative distance ($\vec{r^{*}}$) of particle
emission points in the pair rest frame:
\begin{eqnarray}
\label{cfn}
C(k^{*}) = \langle \left| \Psi^{S}_{- \vec{k^{*}}}
(\vec{r^{*}})\right|^{ 2}\rangle.
\end{eqnarray}
It should be noted that the two particles are generally
produced at non-equal times in their center-of-mass system
and that the wave function in Eq.~(\ref{cfn}) should be
substituted by the Bethe-Salpeter amplitude. The latter depends
on both space $(\vec{r^{*}})$ and time $(t^*)$ separation of the
emission points in the pair rest frame and at small $|t^*|$
coincides with the wave function $\Psi^{S}$
up to a correction ${\cal O}(|t^*/m r^{*2}|)$, where
$m$ is the mass of the lighter particle.
It can be shown that Eq.~(\ref{cfn}) is usually valid better
than to few percent even for particles as light as pions
\cite{Led:90COR,Led:05}.
The wave function
$\Psi^{S}$
represents a stationary solution of the
scattering problem having at large distances
${r}^*$ the asymptotic form of a superposition
of the plane and outgoing spherical waves. It is
approximated by the solution outside
the range of the strong interaction potential taking into
account, at the considered
small k* values, the s-wave part of the scattered wave only:
\begin{eqnarray}
\label{form5}
\Psi_{- \vec{k^{*}}}^{S} (\vec{r^{*}}) \doteq e^{-i\vec{k^{*}} \cdot \vec{r^{*}}}
+ \frac{f^{S}(k^{*})}{r^{*}} e^{ik^{*} \cdot r^{*}},
\end{eqnarray}
with the effective range approximation for the s-wave scattering amplitude:
\begin{eqnarray}
f^{S}(k^{*}) = ( \frac{1}{f^{S}_{0}}
+ \frac{1}{2} d^{S}_{0} k^{*2} - i k^{*} )^{-1},
\end{eqnarray}
where $f^{S}_{0}$ is the scattering length and
$d^{S}_{0}$ is the effective radius for a given total spin S=1 or S=0,
i.e. for a triplet (t) or singlet (s) state respectively. We assume that
particles are produced unpolarized,
i.e.\
$\rho_0=1/4$
of the pairs are in the singlet state and
$\rho_1=3/4$
are in the triplet state.
Then, assuming
a Gaussian distribution in $r^{*}$,
\begin{eqnarray}
d^{3}N/d^{3}r^{*} \sim e^{- \vec{ r^{*} }^{2} /4r_{0}^{2}},
\end{eqnarray}
where $r_{0}$
can be considered as the effective
radius of the source,
the correlation function can be calculated analytically \cite{Led:90COR}:
\begin{eqnarray}
\label{cft}
C(k^*) &=& 1 + \sum_S \rho_S\left[\frac12\left|\frac{f^S(k^*)}{r_0}
\right|^2\left(1-\frac{d_0^S}{2\sqrt\pi r_0}\right)\right. + \nonumber\\
& &\left.\frac{2\Re f^S(k^*)}{\sqrt\pi r_0}F_1(Qr_0)-
\frac{\Im f^S(k^*)}{r_0}F_2(Qr_0)\right] ,
\end{eqnarray}
where
$F_1(z) = \int_0^z dx e^{x^2 - z^2}/z$ and $F_2(z) = (1-e^{-z^2})/z$.
The leading correction
to the correlation function
${\cal O}(|f_0^S|^2 d_0^S/r_0^3)$
is introduced in Eq.~(\ref{cft}) to account for the deviation
of the solution (\ref{form5})
from the true wave function
inside the range of the strong interaction potential.

\subsection{FSI parameters and source sizes}
The $p-\lam$ and $\pbar-\alam$ interaction potentials are relatively
well understood
\cite{Blume:2003,Chung:2003,Wang:9983,Rijnucl:1998,Maessen:1989,Nagels:1979},
which allows us to extract the source radius $r_{0}$ from the fit.
The best fits are compared
with the separate $p-\lam$ and with the
$\pbar-\alam$ correlation functions in
Figure \ref{plaCorrFit} (a), and the combined one in Figure \ref{plaComb}.
The scattering lengths
($f^{s}_{0} = 2.88$ fm, $f^{t}_{0} = 1.66$ fm)
and effective radii ($d^{s}_{0} = 2.92$ fm, $d^{t}_{0} = 3.78$ fm)
from \cite{Wang:9983} have been used
for the $p-\lam$, $\pbar-\alam$ correlation functions.
The systematic errors
on the radius introduced by the uncertainties on the
scattering lengths have
been estimated to be 0.2 fm assuming
spin averaged FSI parameters with $5\%$ uncertainty.
The fit results are summarized in Table \ref{pLaParam}.
The three errors are, from
left to right, the statistical errors and
the systematic
errors introduced by the uncertainty on the purity
correction and
on the scattering length for $p-\lam$ and $\pbar-\alam$ systems
and on the uncertainty in the model for $p-\alam$ and $\pbar-\lam$ systems..
One parameter is free while fitting the $p-\lam$, the $\pbar-\alam$
and the combined correlation functions.
Three parameters are free while  fitting the  $p-\alam$, the $\pbar-\lam$
and the combined correlation functions.
Statistical errors on the radii are larger for
the $p-\lam$, the $\pbar-\alam$
and for the combined correlation functions
than for the corresponding baryon-antibaryon ones.
The $p-\alam$, the $\pbar-\lam$
and the combined correlation functions have a large width
and involve more statistics in the fit of the correlated  $\kStar$-region
as compared with the  $p-\lam$, the $\pbar-\alam$
and the combined correlation functions.

The extracted source radii are close to
the values ($3-4~\mathrm{fm}$) obtained in measurements
performed by the NA49 (SPS) collaboration in Pb+Pb collisions at 158 AGeV \cite{Blume:2003}
and by the E895 (AGS) experiment in Au+Au
collisions at 4, 6, and 8 AGeV \cite{Chung:2003,Led:01Pal}.
This confirms that the particle emitting source size does not change
significantly with beam energy; a result also obtained by studying
two-pion correlations.

The $\pbar-\lam$ and $p-\alam$ scattering lengths have never been measured before.
Hence, they have to be included
as free parameters in the fit to the experimental correlation functions.
In order to limit
the number of free parameters, the following assumptions are made:
i) the spin dependence
is neglected, $f^{s} = f^{t} = f$ and
ii) the effective radius ($d_{0}$) is set to zero.
An extra parameter ${\rm Im} f_0 > 0 $ is added,
taking into account the annihilation channels.

\begin{figure}[hbt!]
\center
\includegraphics[width=9cm, height=7.5cm]{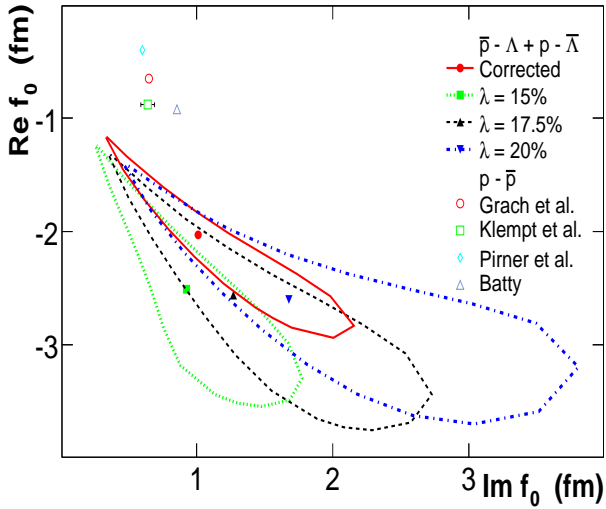}
\caption{(Color online) The combined $(\pbar-\lam)\oplus(p-\alam)$
spin-averaged s-wave scattering length
compared with the previous measurements for the $p-\pbar$
system \cite{Gra:88S609,Bat:89R52,Pir:91S48,Klem:88P368}.
The curves show the one standard deviation contours.Note that for $(\pbar-\lam)\oplus(p-\alam)$ only,
one should read $0.1973 \times {\rm Im} f_0 $ instead of ${\rm Im} f_0$ on the x-axis.}
\label{FitCont}
\end{figure}

The best fits are compared
with the separate $\pbar-\lam$ and $p-\alam$ correlation functions in
Figure \ref{plaCorrFit} (b), and with the combined
one in Figure \ref{plaComb}.
The fitted spin-averaged scattering lengths
for the combined $p-\alam$ and $\pbar-\lam$ systems are compared
with measurements for the $p-\pbar$ system
\cite{Gra:88S609,Klem:88P368,Pir:91S48,Bat:89R52}
in Figure \ref{FitCont}.
The imaginary part of the fitted scattering length is in agreement
with the $p-\pbar$ results while the real part
is more negative.
The error contour represents the statistical errors.
The systematic
error due to the uncertainty on pair purity is investigated by comparing
the best estimated $k^{*}$-dependent
purity with $k^{*}$-independent purity
corrections $\lambda$.
The $k^{*}$-dependent purity
correction tends to decrease the size of the error contour
(the curve labeled ``Corrected") as compared with the one
for the constant purity
(the curve labeled ``$\lambda = 17.5\%$")
and shifts both the real and imaginary parts
of the scattering length in the direction of the $p-\pbar$ values.

The radii extracted from the
fits to the separate and the combined
$p-\alam$
and $\pbar-\lam$ correlation functions
are summarized in Table \ref{pLaParam}.
The errors include from left to right, statistical errors,
systematic errors due to purity,
and systematic errors estimated from varying model parameters.
The error on the radius parameter due to the uncertainties
on the model is estimated to be 0.3 fm.
This error is estimated by fixing
the spin averaged scattering lengths
and by extracting the radius and the effective radius.
For the moment, we do not have any tool to extract
the radius uncertainty related to the neglected p-wave contribution.
A larger radius implies a correlation
over a smaller $k^*$-range
than seen in the data,
which cannot be recovered by increasing
the magnitude of scattering lengths.
However the radii extracted from the $p-\lam$ (and $\pbar-\alam$)
and the $\pbar-\lam$ (and $p-\alam$) are significantly different. The error bars accounting
for all statistic and systematic contributions barely overlap.

\begin{table}[hbt!]
\begin{center}
\begin{tabular}{|c|c|c|c|c|c|}
\hline
Exp. & System & $r_{0}$ (fm)\\ \hline
STAR&   $p-\lam$                                &       $2.97\pm0.34^{+0.19}_{-0.25}\pm0.2$     \\ \hline
STAR&   $\pbar-\alam$                   &       $3.24\pm0.59^{+0.24}_{-0.14}\pm0.2$     \\ \hline
STAR&   $p-\lam\oplus\pbar-\alam$               &       $3.09\pm0.30^{+0.17}_{-0.25}\pm0.2$     \\ \hline
STAR&   $\pbar-\lam$                    &       $1.56\pm0.08^{+0.10}_{-0.14}\pm0.3$     \\ \hline
STAR&   $p-\alam$                       &       $1.41\pm0.10\pm0.11\pm0.3$              \\ \hline
STAR&   $\pbar-\lam\oplus p-\alam$              &       $1.50\pm0.05^{+0.10}_{-0.12}\pm0.3$     \\ \hline
NA49&   $p-\lam$ ($\lambda=0.33$ fixed)         &       $3.8\pm 0.33$                           \\ \hline
NA49&   $p-\lam$ ($\lambda=0.17\pm0.11$ free)&  $2.9\pm0.7$                             \\ \hline
E895&   $p-\lam$ ($\lambda=0.5\pm0.2$ free)     &       $4.5\pm0.7$                             \\ \hline
\end{tabular}
\caption{Comparison of the radius of the source of particles
for $p-\lam$, $\pbar-\alam$, $\pbar-\lam$, $p-\alam$
and combined systems.
For STAR, the three errors are, from
left to right, the statistical errors and
the systematic
errors introduced by the uncertainty on the purity
correction and
on the scattering length for $p-\lam$ and $\pbar-\alam$ systems
and on the uncertainty in the model for $p-\alam$ and $\pbar-\lam$ systems.
For NA49 \cite{Blume:2003} and E895 \cite{Led:01Pal},
the $\lambda$ parameter
represents the pair purity.}
\label{pLaParam}
\end{center}
\end{table}

\section{DISCUSSION}
\begin{figure}[hbt!]
\center
\includegraphics[width=9cm, height=7.5cm]{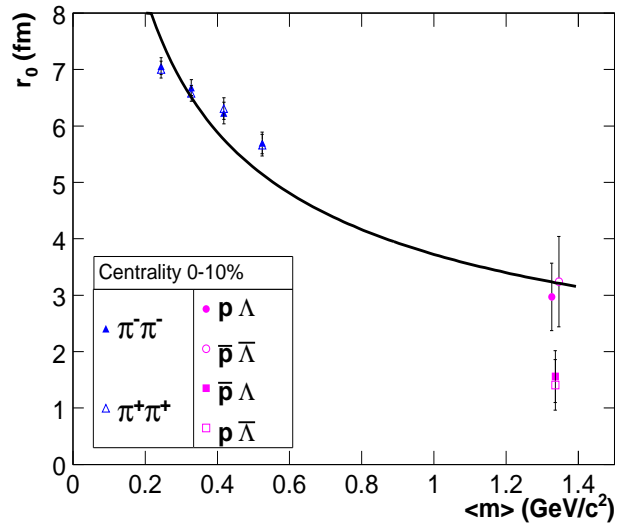}
\caption{(Color online) Pion source size \cite{Adams:2004e}
compared with proton and lambda
source sizes as a function of the mean
transverse mass ($\langle m_{t} \rangle$).
The curve shows the $\langle m_t \rangle^{-1/2}$
dependence with an arbitrary normalization.}
\label{RinvvsMt}
\end{figure}

The difference in radii between $p-\lam$ ($\pbar-\alam$)
and $\pbar-\lam$
($p-\alam$) is unexpected. Indeed, it would
imply a novel dynamical
space-momentum correlation between proton ($\pbar$)
and $\alam$ ($\lam$).
Strong space-momentum correlations are exhibited
in Au+Au collisions at RHIC.
These are understood to arise from the collective flow
of massive particles \cite{Adams:2004e,NONID:2003}.
This effect, however, would not lead to a difference
between the source size measured from $p-\lam$
and $\pbar-\lam$ correlations.
In Figure 4, the source sizes from proton-Lambda correlations
and pion-pion correlations are plotted as a function of the mean
of the particles' transverse masses.
The decrease of the source size with increasing mean transverse mass
is in qualitative agreement with expectations
from collective flow \cite{BW:2003}.
The curve in Fig.\ref{RinvvsMt}  represents an arbitrarily
normalized $\langle m_t \rangle^{-1/2}$ dependence.
This dependence is expected within some hydrodynamics-motivated
models \cite{Toma:00N663}.
The data are in reasonable agreement with this expectation.
In addition, a possible difference between radii pointed by data
may imply that baryon - antibaryon pairs are produced
close in space, a dynamic correlation that is not in
baryon - baryon pairs.

While a novel space-momentum correlation between proton and $\alam$ cannot
be ruled out, the difference between the radii extracted from $p-\lam$ and
$\pbar-\lam$ correlation functions may come from an imperfect treatment of
the purity correction. Indeed, we have assumed that any pairs
that are not composed of two primary particles are not correlated.
However,
Table \ref{TabResCorr} shows that a number of such pairs may carry a
residual correlation from their parents \cite{Stav}. For example, the interaction between
a primary proton and a $\Sigma^0$ may not be completely washed out when
constructing the $p-\lam$ correlation function with the $\lam$ being the
$\Sigma^0$ daughter as the $\lam$ carries most of the momentum of its parent.
This effect was found to be on the order of $10\%$ \cite{WangPRC}.
However, none
of the interactions between the pairs listed in Table \ref{TabResCorr}
are known. We are thus unable to perform any reliable correction or
error estimate. At that stage, we show the $p-\alam$ and $\pbar-\lam$
correlation function corrected with the best estimate of the purity
assuming no residual correlations. We extract source radii and
scattering length parameters acknowledging that the values may be
biased by the presence of residual correlations.

\begin{table}[hbt!]
\begin{tabular}{|c|c|c|c|}
 \hline
Pairs & Fractions \\ \hline
$p_{prim}-\lam_{prim}$                  & $15\%$ \\ \hline
$p_{\lam}-\lam_{prim}$                  & $10\%$ \\ \hline
$p_{\sigp}-\lam_{prim}$                 & $ 3\%$ \\ \hline
$p_{prim}-\lam_{\sigz}$                 & $11\%$ \\ \hline
$p_{\lam}-\lam_{\sigz}$                 & $ 7\%$ \\ \hline
$p_{\sigp}-\lam_{\sigz}$                & $ 2\%$ \\ \hline
$p_{prim}-\lam_{\Xi}$                   & $ 9\%$ \\ \hline
$p_{\lam}-\lam_{\Xi}$                   & $ 5\%$ \\ \hline
$p_{\sigp}-\lam_{\Xi}$                  & $ 2\%$ \\ \hline
$p_{prim}-p_{prim}$                     & $ 7\%$ \\ \hline
\end{tabular}
\caption{Summary of the main fractions of pairs
containing particles from particle decays included
in $p-\lam$, $p-\alam$, $\pbar-\lam$, $\pbar-\alam$
correlation functions assuming the absence of residual correlations.
$\lam_{\Xi}$, are $\lam$ ($\alam$) decay products
of $\xim, \xiz$ ($\axim, \axiz$),
$\lam_{\sigz}$, are $\lam$ ($\alam$) decay products
of $\sigz$ ($\asigz$),
$p_{\lam}$ are $p$ ($\pbar$) decay products
of $\lam$ ($\alam$),
$p_{\sigp}$ are $p$ ($\pbar$) decay products
of $\sigp$ ($\asigp$),
$\lam_{prim}$ and $p_{prim}$ represent primary
$\lam$ ($\alam$) and $p$ ($\pbar$).
The remaining $29\%$ represents misidentified
$p$ ($\pbar$) and
reconstructed fake $\lam$ ($\alam$).}
\label{TabResCorr}
\end{table}

\section{CONCLUSION}
Constructing $p-\lam$, $p-\alam$, $\pbar-\lam$,
$\pbar-\alam$, we have gathered information about the
space-time features of baryon and antibaryon emission
and about the interaction
in $p-\alam$ and $\pbar-\lam$ systems.
The source radii extracted from $p-\lam$ and $\pbar-\alam$
corrrelation function agree with the flow expectation.
The radii extracted from $\pbar-\lam$ and $p-\alam$
are significantly smaller.
Final-state interactions parameters,
such as spin averaged s-wave
scattering length, have been
extracted from $\pbar-\lam$ and $p-\alam$
correlation functions. The real part
of the scattering length
appears to be negative while
the imaginary part is positive, the
latter being required by the unitarity
due to the open annihilation
channels.
These results demonstrate that
correlation measurements can be used to
study the two-particle strong interaction
for particle combinations
that are difficult to access by other means,
including traditional scattering
experiments.

\begin{acknowledgments}
We thank the RHIC Operations Group and RCF at BNL, and the
NERSC Center at LBNL for their support. This work was supported
in part by the HENP Divisions of the Office of Science of the U.S.
DOE; the U.S. NSF; the BMBF of Germany; IN2P3, RA, RPL, and
EMN of France; EPSRC of the United Kingdom; FAPESP of Brazil;
the Russian Ministry of Science and Technology; the Ministry of
Education and the NNSFC of China; IRP and GA of the Czech Republic,
FOM of the Netherlands, DAE, DST, and CSIR of the Government
of India; Swiss NSF; the Polish State Committee for Scientific
Research; STAA of Slovakia, and the Korea Sci. \& Eng. Foundation.

\end{acknowledgments}

\bibliography{pLaPaper}

\end{document}

%% file: STAR_authors.tex
\affiliation{Argonne National Laboratory, Argonne, Illinois 60439}
\affiliation{University of Bern, 3012 Bern, Switzerland}
\affiliation{University of Birmingham, Birmingham, United Kingdom}
\affiliation{Brookhaven National Laboratory, Upton, New York 11973}
\affiliation{California Institute of Technology, Pasadena, California 91125}
\affiliation{University of California, Berkeley, California 94720}
\affiliation{University of California, Davis, California 95616}
\affiliation{University of California, Los Angeles, California 90095}
\affiliation{Carnegie Mellon University, Pittsburgh, Pennsylvania 15213}
\affiliation{Creighton University, Omaha, Nebraska 68178}
\affiliation{Nuclear Physics Institute AS CR, 250 68 \v{R}e\v{z}/Prague, Czech Republic}
\affiliation{Laboratory for High Energy (JINR), Dubna, Russia}
\affiliation{Particle Physics Laboratory (JINR), Dubna, Russia}
\affiliation{University of Frankfurt, Frankfurt, Germany}
\affiliation{Institute of Physics, Bhubaneswar 751005, India}
\affiliation{Indian Institute of Technology, Mumbai, India}
\affiliation{Indiana University, Bloomington, Indiana 47408}
\affiliation{Institut de Recherches Subatomiques, Strasbourg, France}
\affiliation{University of Jammu, Jammu 180001, India}
\affiliation{Kent State University, Kent, Ohio 44242}
\affiliation{Lawrence Berkeley National Laboratory, Berkeley, California 94720}
\affiliation{Massachusetts Institute of Technology, Cambridge, MA 02139-4307}
\affiliation{Max-Planck-Institut f\"ur Physik, Munich, Germany}
\affiliation{Michigan State University, East Lansing, Michigan 48824}
\affiliation{Moscow Engineering Physics Institute, Moscow Russia}
\affiliation{City College of New York, New York City, New York 10031}
\affiliation{NIKHEF and Utrecht University, Amsterdam, The Netherlands}
\affiliation{Ohio State University, Columbus, Ohio 43210}
\affiliation{Panjab University, Chandigarh 160014, India}
\affiliation{Pennsylvania State University, University Park, Pennsylvania 16802}
\affiliation{Institute of High Energy Physics, Protvino, Russia}
\affiliation{Purdue University, West Lafayette, Indiana 47907}
\affiliation{Pusan National University, Pusan, Republic of Korea}
\affiliation{University of Rajasthan, Jaipur 302004, India}
\affiliation{Rice University, Houston, Texas 77251}
\affiliation{Universidade de Sao Paulo, Sao Paulo, Brazil}
\affiliation{University of Science \& Technology of China, Anhui 230027, China}
\affiliation{Shanghai Institute of Applied Physics, Shanghai 201800, China}
\affiliation{SUBATECH, Nantes, France}
\affiliation{Texas A\&M University, College Station, Texas 77843}
\affiliation{University of Texas, Austin, Texas 78712}
\affiliation{Tsinghua University, Beijing 100084, China}
\affiliation{Valparaiso University, Valparaiso, Indiana 46383}
\affiliation{Variable Energy Cyclotron Centre, Kolkata 700064, India}
\affiliation{Warsaw University of Technology, Warsaw, Poland}
\affiliation{University of Washington, Seattle, Washington 98195}
\affiliation{Wayne State University, Detroit, Michigan 48201}
\affiliation{Institute of Particle Physics, CCNU (HZNU), Wuhan 430079, China}
\affiliation{Yale University, New Haven, Connecticut 06520}
\affiliation{University of Zagreb, Zagreb, HR-10002, Croatia}

\author{J.~Adams}\affiliation{University of Birmingham, Birmingham, United Kingdom}
\author{M.M.~Aggarwal}\affiliation{Panjab University, Chandigarh 160014, India}
\author{Z.~Ahammed}\affiliation{Variable Energy Cyclotron Centre, Kolkata 700064, India}
\author{J.~Amonett}\affiliation{Kent State University, Kent, Ohio 44242}
\author{B.D.~Anderson}\affiliation{Kent State University, Kent, Ohio 44242}
\author{D.~Arkhipkin}\affiliation{Particle Physics Laboratory (JINR), Dubna, Russia}
\author{G.S.~Averichev}\affiliation{Laboratory for High Energy (JINR), Dubna, Russia}
\author{S.K.~Badyal}\affiliation{University of Jammu, Jammu 180001, India}
\author{Y.~Bai}\affiliation{NIKHEF and Utrecht University, Amsterdam, The Netherlands}
\author{J.~Balewski}\affiliation{Indiana University, Bloomington, Indiana 47408}
% ...

\collaboration{STAR Collaboration}\noaffiliation